\documentclass[conference]{IEEEtran}
\IEEEoverridecommandlockouts
\usepackage[T1]{fontenc}
\usepackage{graphicx}
\usepackage{color}
\usepackage[colorlinks]{hyperref}
\hypersetup{
	linkcolor={blue}
}
\usepackage{todonotes}
\usepackage{orcidlink}
\usepackage{tikz}
\usetikzlibrary{shapes.geometric, arrows}

\tikzstyle{startstop} = [rectangle, minimum width=5cm, minimum height=1cm, text centered, draw=black, fill=red!30, rotate=90]
\tikzstyle{process} = [rectangle, minimum width=5cm, minimum height=1cm, text centered, draw=black, fill=orange!30, rotate=90]
\tikzstyle{quantization} = [rectangle, minimum width=5cm, minimum height=1cm, text centered, draw=black, fill=yellow!30, rotate=90]
\tikzstyle{arrow} = [thick,->,>=stealth]

\begin{document}
\title{Architecture, Simulation and Software Stack to Support Post-CMOS Accelerators: The ARCHYTAS Project
\thanks{\textbf{Acknowledgement}. This work has been funded by the European Union via the European Defence Fund project ARCHYTAS under grant agreement nr.101167870. 
Views and opinions expressed are however those of the author(s) only and do not necessarily reflect those of the European Union or the European Commission. 
Neither the European Union nor the granting authority can be held responsible for them.}
}

%\titlerunning{The ARCHYTAS Project}
\author{
\IEEEauthorblockN{Giovanni Agosta\IEEEauthorrefmark{1}\orcidlink{0000-0002-0255-4475},
Stefano Cherubin\IEEEauthorrefmark{6}\orcidlink{0000-0002-5579-5942},
Derek Christ\IEEEauthorrefmark{5}\orcidlink{0009-0005-4234-6362},
Francesco Conti\IEEEauthorrefmark{3}\orcidlink{0000-0002-7924-933X},
Asbj{\o}rn Djupdal\IEEEauthorrefmark{6}\orcidlink{0009-0002-3571-2828},\\
Matthias Jung\IEEEauthorrefmark{5}\IEEEauthorrefmark{7}\orcidlink{0000-0003-0036-2143},
Georgios Keramidas\IEEEauthorrefmark{4}\orcidlink{0000-0003-0460-6061},
Roberto Passerone \IEEEauthorrefmark{2}\orcidlink{0000-0001-6315-1023},
Paolo Rech \IEEEauthorrefmark{2}\orcidlink{0000-0002-0821-1879},
Elisa Ricci \IEEEauthorrefmark{2}\orcidlink{0000-0002-0228-1147},\\
Philippe Velha\IEEEauthorrefmark{2}\orcidlink{0000-0002-1826-2370},
Flavio Vella\IEEEauthorrefmark{2}\orcidlink{0000-0002-7924-933X},
Kasim Sinan Yildirim \IEEEauthorrefmark{2}\orcidlink{0000-0002-9528-6923},
Nils Wilbert\IEEEauthorrefmark{5}\orcidlink{0009-0005-1108-462X} 
}
%
%\authorrunning{G. Agosta et al.}
\IEEEauthorblockA{\IEEEauthorrefmark{1}\href{https://deib.polimi.it}{DEIB -- Politecnico di Milano}, Italy {name.surname@polimi.it} \\
\IEEEauthorrefmark{2}\href{https://unitn.it}{Università degli Studi di Trento}, Italy {name.surname@unitn.it}\\
\IEEEauthorrefmark{3}\href{https://unibo.it}{Alma Mater Università degli Studi Bologna}, Italy {name.surname@unibo.it}\\
\IEEEauthorrefmark{4} \href{https://www.auth.gr/}{Aristotle University of Thessaloniki}, Greece {name.surname@auth.gr}\\
\IEEEauthorrefmark{5} \href{https://www.uni-wuerzburg.de/}{Julius-Maximilians Universit\"at W\"urzburg}, Germany\\ {\{derek.christ, m.jung, nils.wilbert\}@uni-wuerzburg.de}\\
\IEEEauthorrefmark{6}\href{https://ntnu.no}{NTNU -- Norwegian University of Science and Technology}, Norway {\{stefano.cherubin,djupdal\}@ntnu.no}\\
\IEEEauthorrefmark{7}\href{https://iese.fraunhofer.de}{Fraunhofer IESE}, Germany {matthias.jung@iese.fraunhofer.de}
}
}
\IEEEoverridecommandlockouts
%\IEEEpubid{\makebox[\columnwidth]{979-8-3315-3477-6\/25\/\$31.00~\copyright2025 IEEE \hfill}
%\hspace{\columnsep}\makebox[\columnwidth]{ }}

\maketitle              % typeset the header of the contribution
%\IEEEpubidadjcol

\begin{abstract}
ARCHYTAS aims to design and evaluate non-conventional hardware accelerators, in particular, optoelectronic, volatile and non-volatile processing-in-memory, and neuromorphic, to tackle the power, efficiency, and scalability bottlenecks of AI with an emphasis on defense use cases (e.g., autonomous vehicles, surveillance drones, maritime and space platforms).
In this paper, we present the system architecture and software stack that ARCHYTAS will develop to integrate and support those accelerators, as well as the simulation software needed for early prototyping of the full system and its components.
\end{abstract}


\section{Introduction}
\label{sec:intro}
Artificial Intelligence (AI) is becoming pervasive in many embedded applications, including the development of control systems for autonomous vehicles (uncrewed aerospace, terrestrial, and marine vehicles, or UxV).
AI has the potential to enhance the efficiency and accuracy of such systems across the entire space of their operation, from computer vision tasks necessary to detect obstacles in a timely manner to advanced decision making in potentially dangerous conditions~\cite{mcenroe2022survey,ma2020ai}.
Current AI algorithms, however, have significant computational requirements, which may reflect into energy consumption that is incompatible with the requirements of UxV, particularly unmanned aerial vehicles (UAV).
Although other types of UxV may be endowed with more abundant energy resources, high energy consumption also implies the need for more complex thermal dissipation systems, such as fans, which may reduce the overall reliability of the system -- an undesirable effect in, e.g., uncrewed vehicles employed in search and rescue.

Increase in energy consumption is particularly significant for AI algorithms that run on traditional processors, such as computer vision applications that require AI models based on Deep Neural Networks (DNNs), with large number of parameters and complex structures. 
In embedded systems and edge computing devices, such as drones or remote sensors, high-energy requirements can be a critical limitation. 
These devices typically have limited power resources and need to operate for extended periods in the field without recharging. 
Thus, running power-hungry AI algorithms on these devices can rapidly deplete their energy reserves and reduce their operational time and effectiveness.
Energy consumption also translates into heat that must be dissipated efficiently to avoid thermal-related component aging~\cite{fornaciari2022textarossa}.
Once more, this aspect is complicated and aggravated by the confined environment of on-board and embedded systems.

To overcome these challenges, there is a growing need for specialized hardware that can accelerate AI workloads~\cite{aaleskog2022recent}. 
These accelerators are designed to provide more efficient execution of AI algorithms compared to traditional general-purpose CPUs or GPUs, enhancing the performance and efficiency of AI algorithms, enabling faster, energy-efficient, and more accurate processing of vast amounts of data. 
Furthermore, high reliability and robust security are also fundamental design considerations for the AI accelerators that would need to withstand extreme environments.
With the end of Moore's law and Dennard scaling, research on new materials and computing paradigms suggests designing novel architectures based on more efficient devices that unlock significantly enhanced performance-per-Watt~\cite{theis2017end,kim2024future}. 
At the same time, the limitation of the Von Neumann model in computing data-centric workloads requires the rethinking of micro-architectures, the exploration of newer compute paradigms and Instruction Set Architectures (ISA) to reduce latency, and to improve performance-per-Watt efficiency and programming productivity~\cite{sebastian2020memory}.

The ARCHYTAS project aims to investigate and study the feasibility of non-conventional AI accelerators for defence applications that take advantage of novel technologies at the device and package level: optoelectronic-based accelerators, volatile and non-volatile processing-in-memory, and neuromorphic devices. The project will also investigate the integration of CMOS-based systems with analogue accelerator devices and their organisation by integrating them in a multi-chip (chiplet) configuration. 
Moreover, ARCHYTAS is set to pioneer new programming models aiming to boost programmability, performance portability, and overall productivity of the latest emerging parallel systems, all through a groundbreaking HW-AI co-design.

\paragraph{The ARCHYTAS Consortium}
is coordinated by Iveco Defense Vehicles (IDV), a leading European player in the defense land vehicles market, with Sener Aerospacial, Maritime Robotics, Ubotica, and Nurenda providing use case scenarios. STMicroelectronics, Upmem, Instrumentation Technologies, and Brightelligence provide the technical backbone from the industrial perspective, while the research is supported by a strong selection of European research organizations, including the Universities of Trento, Bologna, Rennes, Wurzburg, Groningen, Thessaloniki, Politecnico di Milano, Sorbonne Université, and NTNU, as well as the Fraunhofer, Namlab and CSIC reasearch centers. Spinverse, an innovation consultancy firm, supports the project management.

\section{Post-CMOS Acceleration}
\label{sec:accel}

CMOS technology has evolved tremendously over the last 50 years \cite{etiemble2018,Shalf2020}
However, the Von Neumann model \cite{hall2013} has remained the fundamental principle that links all forms of computing from the early 50's to the latest processors.

A central processing unit (\textbf{CPU}) goes through a set of registers that contain instructions and data to compute and accordingly interacts synchronously with memory units and with all peripherals through different types of I/O interfaces.
An early metric to assess performance has been the clock frequency (\textbf{$f_{clk}$}), the faster the CPU can compute, the higher the performance. However, this one step-at-a-time approach has some wider repercussion: the so-called Von Neuman Bottleneck. At each step, data needs to be accessed fast enough and computed to provide a result. This constraint poses a fundamental limit on compute-driven system: data movement.

Over the years, several mitigation strategies have been employed (cache hierarchy~\cite{hall2013}, modified Harvard architecture \cite{Ravi1989}, branch prediction~\cite{Paschos2018}, pipe-lining, ...) but the fundamental one-step-at-a-time concept remains unchanged hindering each generation of CPU.

Until 2005~\cite{Torrellas2016,tadonki2020}, clock frequencies have been increasing until hitting the ``thermal wall''.
Basically, all synchronous systems are based on the switching of transistor that transfer charges. It can be shown that the power loss is proportional to $C_b V^2 f_{clk}$ where $C_b$ is the amount of charge that is moving around to represent a bit, and $V$ is the voltage supply.
That simple relation provides a practical limit to $f_{clk}$ of around 5GHz related to thermal management and power consumption.
After that, performance has been improved mainly through parallelization~\cite{venu2011}, at the cost of more complex and challenging data movement management (data coherence and consistency \cite{Torrellas2016}, shared memory and cache hierarchy complexity, etc.) and an impact on the software tool-chain stack\cite{Hill2014}.
With the advent of GPUs offering massive parallelization, those challenges elevated to an even-higher level workload optimization, which is a hot topic of research~\cite{Nugteren2014}.

With ``better defined'' workloads like Machine Learning and AI training, GPUs have found their flagship application, in particular, in HPC.
Nonetheless, the conceptual flaw remains and HPC systems remain bandwidth-bound~\cite{Williams2009,kaplan2020} instead of compute bound.
For instance, AI training~\cite{erdil2024} suffers from a limited utilization rate (sometimes below 50\%) hinting that a deep understanding of hardware practical limits is crucial for AI future development.
The additional obstacle that has prompted a global energy race~\cite{Energy2025} is the power demand from data centers, which is becoming one of the predominant users of global energy.
To overcome these limits, we endeavor to explore a change of paradigm that drops the Von Neumann model. Since the main bottleneck originates from data, we foresee a switch from compute-driven systems to data-driven systems. In other words, instead of moving data to the compute unit, we propose to bring the computation to the data.
In ARCHYTAS, we declined this ambition in different forms:
\begin{itemize}
    \item Processing-In-Memory \cite{Kim2020,vincon2019,Andrighetti2020,ghose2018}(\textbf{PIM}) where processing takes place within the memory elements themselves, by exploiting their physical characteristics.
    \item Processing-Near-Memory \cite{Kim2017,SINGH2019} (\textbf{PNM}) where processing takes place on compute units 2.5D/3D-stacked over the memory elements.
    \item Processing-On-the-Flight (\textbf{POF}) where data is processed as it travels. Since most data in an HPC are transported in the optical domain, we propose to explore photonic accelerators (\textbf{PA}) 
    
\end{itemize} 

The relentless pursuit of energy efficiency and enhanced computational capabilities has propelled the evolution of computing systems towards greater parallelism and heterogeneity~\cite{Taylor2012} requiring new solutions. 
For the photonic accelerator, we propose to develop a framework that allows space exploration of different proposed implementations inside a computing system. Many demonstrator have been realized however, the presented work usually focuses on a restricted number of metrics like throughput, power consumption, or latency. Rarely is a system approach presented and generally lacks several layers of the full stack. Our scope is to fully address this gap and offer the community a clear set of tools to benchmark and test over the full stack (from hardware to software) accelerators inserted with more classical computing nodes.

\section{System Architecture for Accelerator Integration}
\label{sec:arch}
To ``extract'' all the potential performance and energy efficiency gains of post-CMOS accelerators based on the PIM, PNM, POF paradigms, it is necessary to fully integrate them into a system architecture that 1) feeds accelerators with sufficient data; 2) connects and combines them so that they can be used cooperatively on a single target task and 3) ultimately, allows for the deployment of complex, multi-layered AI algorithms such as DNNs and Vision Transformers.

To achieve this goal, the system and architecture work in the ARCHYTAS project encompasses all activities related to the design of ``conventional'' digital subsystems that accompany and complement the post-CMOS accelerators proposed in the project, as well as the exploration of the design of the full accelerator system.
The approach that ARCHYTAS will pursue is that of designing a \textbf{Scalable Compute Fabric}, i.e., a compute architecture that can be tailored to a diversity of sizes and can integrate heterogeneous accelerators (both post-CMOS and digital) depending on target goals and constraints.

The ARCHYTAS Scalable Compute Fabric will include both general-purpose processors (GPPs, in particular based on RISC-V) and a scalable heterogeneous architecture coupling digital neural processing units (NPUs), multi-core digital signal processors (DSPs), both programmable and non-programmable, and non-conventional units, such as neural accelerators based on non-volatile memory (NVM).
To enable the integration of multiple highly heterogeneous compute blocks, ARCHYTAS will focus on a scalable, tiled architecture based on a Network-on-Chip (NoC) to enable performance up-scaling, while at the same time enabling high-bandwidth utilization of available compute resources with high bandwidth (such as High Bandwidth Memories).

\begin{figure*}[t]
    \centering
    \includegraphics[width=0.78\linewidth]{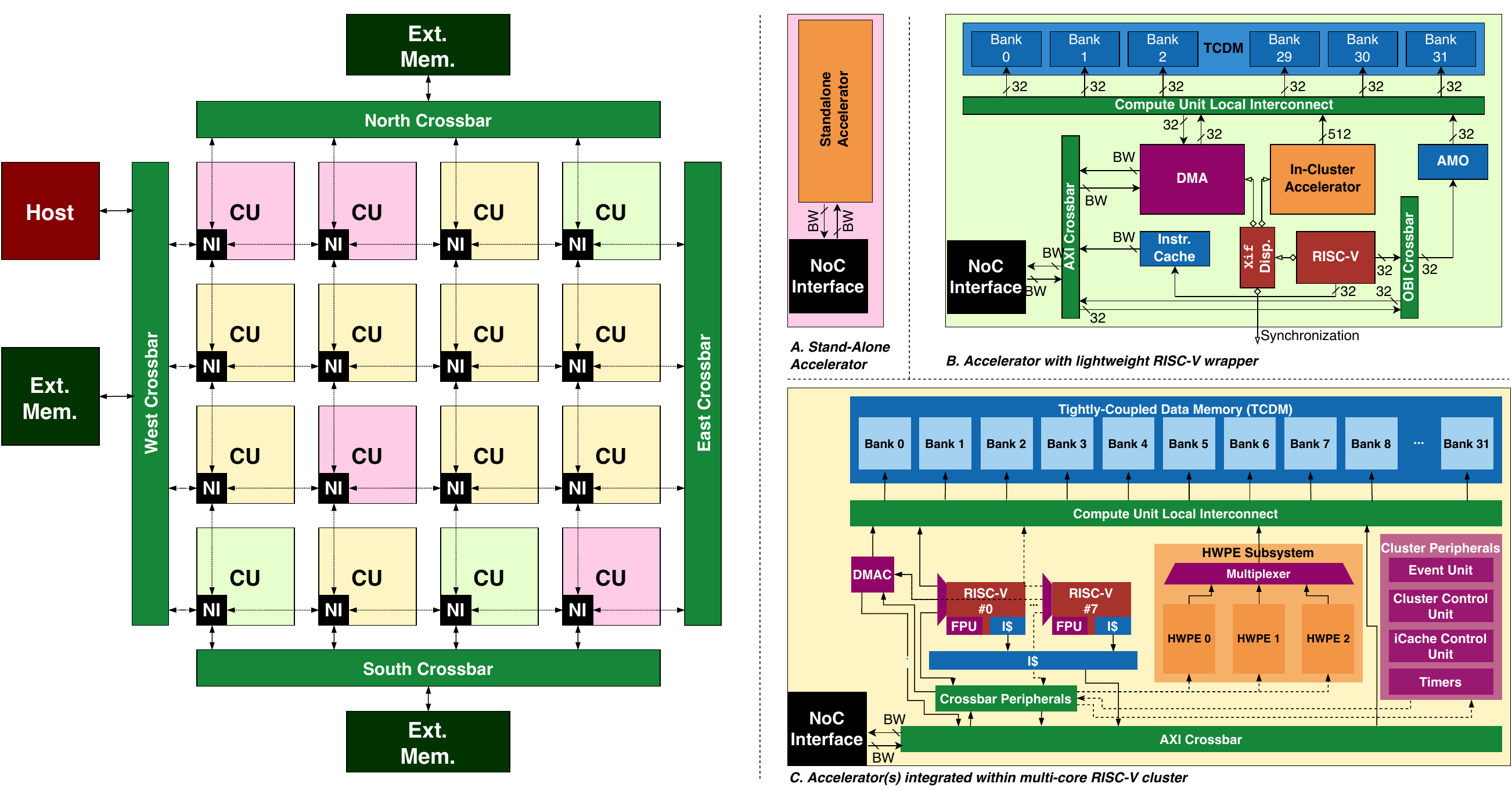}
    \caption{ARCHYTAS Scalable Compute Fabric (left) with three supported Compute Unit (CU) templates: \textit{A.} stand-alone accelerator with NoC interface; \textit{B.} accelerator integrated within a light-weight wrapper with RISC-V controller, local tightly-coupled data memory, and DMA; \textit{C.} accelerator(s) integrated in a multi-core RISC-V cluster based on the PULP platform paradigm.}
    \label{fig:scalable_fabric}
\end{figure*}

The Scalable Compute Fabric, which is shown in Fig.~\ref{fig:scalable_fabric}, will employ multiple levels of heterogeneity.
On the one hand, each tile connected to the NoC -- called a \textbf{Compute Unit} (CU) -- will potentially different from other tiles.
On the other hand, the tiles will adhere to one of several internal microarchitectural paradigms, further increasing heterogeneity: for example, each CU could be based on a stand-alone accelerator provided as a ``black box'' and exposing a high-level NoC interface; or it could be an internally complex unit including also cores, digital accelerators/neural processing units, and/or SRAM-based Digital In-Memory Computing devices.
In this latter category, as exemplified in Fig.~\ref{fig:scalable_fabric}, we plan to support either a lightweight wrapper around an accelerator, to provide local memory, DMA, and a RISC-V based controller core; or a complete cluster based on the PULP platform heterogeneous paradigm~\cite{conti2023marsellus,conti2024opensourceheterogeneoussocsai}.
We will base our work on the open-source FlooNoC~\cite{fischer2025floonoc} Network-on-Chip infrastructure.
Moreover, the project will offer microarchitectural support for techniques such as tensor sparsification to maximize the utilization of compute units while minimizing communication overheads on highly sparse data.

Our methodology will be based on full microarchitecture design and characterization of single subsystems and small fabrics of tiles, estimating area and energy efficiency of the system.
ARCHYTAS will also study how to up-scale such heterogeneous architectures to achieve the target performance and energy efficiency by means of suitable Network-on-Chip (NoC) infrastructure, also using high-level simulation~\cite{bruschi2021gvsoc} \& FPGA emulation frameworks to study the deployment of larger workloads on the composite heterogeneous compute fabric.

To enable high-throughput and full-efficiency usage of accelerators, emphasis will be put on further development of the intra-chip NoC; innovative NoC topologies will be studied to achieve a tunable balance between cost and performance for chips that boast hundreds to thousands of highly heterogeneous processing units. 
ARCHYTAS will define a toolchain incorporating approximate NoC floor-planning and link routing to provide rapid yet precise cost and performance estimations, and will develop methods for efficient Design space exploration to find optimal architectures, using both Mixed-Integer Linear Programming (MILP), following the approach in ArchEx, and Boolean techniques, such as Satisfiability Modulo Theory (SMT) for efficient exploration of the design space. The final goal is to reduce chip area and power consumption, thus we will adopt Low-Radix Topologies and Design for Routability design principle. System-level simulation will also be introduced using
an iterative optimisation approach to speed up the execution and deduce constraints to guide the solver to the optimal solution more quickly. The toolchain will not only support NoC topology selection, but also packaging choice and solutions such as chiplet integration.

\section{Simulation of Processing-In-Memory with DRAMSYS}
\label{sec:dramsys}
The inclusion of non-conventional components requires new methods for system level simulations. DRAMSys~\cite{jung2015dramsys} will be employed in ARCHYTAS to investigate Processing-in-Memory (PIM) architectures within both DRAM and non-volatile memories (NVM). Consequently, ARCHYTAS aims to augment the DRAMSys tool with PIM and NVM functionalities. This enhancement will enable the exploration of PIM design possibilities and establish an experimental platform for hardware/software co-design based on PIM technology. DRAMSys4.0~\cite{steiner2022dramsys4} is a flexible, fast, and open-source DRAM subsystem design space exploration framework based on SystemC TLM-2.0. This simulator provides highly accurate simulations of memory systems at a remarkably fast simulation speed. In ARCHYTAS, the DRAMSys tool shall be enhanced by PIM capabilities to explore the PIM design space and provide an experimental platform for PIM-based hardware/software co-design~\cite{christ2024pimsys}. 
For system-level simulations, the PULP GVSoC~\cite{bruschi2021gvsoc} simulator will be adapted to the architecture developed in the System- and Micro-Architecture task and cooperating on its integration with DRAMSys.

\section{Compiler Support for Neural Network Mapping}
\label{sec:raptor}
Compiler technology plays a critical role in ARCHYTAS, as the mediator between application needs and the hardware (and its abstraction layer).
The key goal of the ARCHYTAS compiler and runtime system is to support the mapping of AI computationally and/or memory intensive kernels to the accelerators. 
This entails the conversion from the neural network description format down to the machine code~\cite{xing2019depth,li2020deep}, as well as to the non-Von Neumann accelerators~\cite{zhao2018hardware,shin2023pimflow}.

\subsection{Compilers for optimized code generation}

Accelerating computationally intensive components of deep neural network (DNN) models remains a complex and non-trivial challenge, particularly due to the involvement of intricate non-linear activation functions. The optimization process must simultaneously account for numerous parameters, including kernel configurations and the characteristics of the target hardware architecture. Moreover, existing modern compilers often fall short in generating highly optimized code, necessitating the use of manually vectorized routines (e.g., leveraging SSE or AVX intrinsics) to achieve acceptable performance~\cite{anderson2015automatic}. The ARCHYTAS project aims to address these challenges by developing an automated toolchain that steers the compilation process while considering the presence and utilization of scratchpad memories in the system architecture~\cite{narasimhan2021practical}.

\subsection{Pruning and sparsification for digital and analogue devices}

The main performance bottleneck in many data-centric applications comes from data movements, primarily due to the inherent limitations of Von Neumann architectures. These architectures operate by repeatedly transferring data and instructions between memory and processing units — a process that becomes particularly inefficient in data-intensive applications. This architectural paradigm severely limits the scalability of AI and Machine Learning workloads, especially during inference, where high throughput, low latency, and energy efficiency are essential~\cite{gholami2024ai}. 
This is particularly evident in complex models such as transformer-based architectures~\cite{vaswani2017attention}, and especially in Vision Transformers (ViTs)~\cite{vasu2024mobileclip}, where deploying such computationally intensive models on edge devices remains an open challenge. 
The challenge is further amplified in resource-constrained environments, where architectural efficiency must be balanced with stringent trade-offs in accuracy, performance, energy, and latency. 

To address these issues, ARCHYTAS focuses on a paradigm shift: keeping the computation close to where the data resides, leveraging Processing-In-Memory (PIM) to eliminate or reduce data movement. However, even with such architectural improvements, further algorithmic and structural optimizations are critical to maximize the benefits of near-data processing.
In this context, \emph{sparsification}, \emph{pruning}, and \emph{dynamic quantization} become essential techniques:
\begin{itemize}
    \item \textbf{Sparsification} refers more broadly to techniques that induce or exploit sparsity in neural networks. This includes both unstructured (fine-grained) and structured (block-wise or filter-level) approaches, applied to weights, neurons, gradients, and activations. Sparsification not only reduces computational and memory overhead but can also improve generalization and robustness~\cite{hoefler2021sparsity}. Importantly, sparsity can be introduced during training, post-training, or even as part of the model initialization.
    
    \item \textbf{Pruning} is a specific form of sparsification that systematically removes redundant or non-informative weights, typically after training. This reduces model size and computation, minimizing data traffic within or outside the memory hierarchy and enabling more efficient execution on resource-constrained systems.

    \item \textbf{Dynamic quantization} reduces the precision of model weights and activations during inference, often down to INT8 or lower, without significant loss in accuracy. This approach shrinks memory footprint and bandwidth requirements while accelerating inference on low-precision digital and mixed-signal hardware platforms~\cite{gholami2022survey,liu2021post}.
\end{itemize}

These techniques are particularly effective when integrated at both the system and architectural levels, and further leveraged through compiler intrinsics and system software support. For example, in optoelectronic-based accelerators, post-training optimization can be implemented within the analog signal representation of the photonic engine, enabling model compression techniques such as pruning and quantization to be embedded directly in the optical domain~\cite{shen2017deep,lin2022all}. This approach aligns tightly with the constraints and capabilities of non-conventional accelerators, such as those being developed within the ARCHYTAS project. Recent research has demonstrated that combining sparsification and quantization with optical computing not only reduces the memory footprint and energy usage but also enables efficient in-situ inference with low latency~\cite{xu202111}. This co-design philosophy—linking algorithmic optimization with photonic hardware primitives—is increasingly seen as essential for achieving scalable and energy-efficient AI systems~\cite{feldmann2021parallel}.

%% introducing quantization and sparsificaiotn. 

% Given the wide differences among the PIM platforms, the goal must be to optimise the network after training – this is challenging, as we lose some degrees of freedom that are available with in-training sparsification and pruning. However, we can perform the transformation entirely within the compiler. To optimise the deployed network, we will leverage a combination of posteriori pruning and sparsification with the following techniques. 

\subsection{Precision tuning approach}
ARCHYTAS will explore the non-conventional programming paradigm of approximate computing~\cite{bosio2022approximate,stanley2020exploiting} in combination with non-conventional computer architectures, such as in-memory processing and neuromorphic computing. 

Error-tolerating applications are increasingly common, particularly among AI-based applications. Proposals have been made at the hardware level to take advantage of inherent perceptual limitations, redundant data, or reduced precision input~\cite{venkatesan2011macaco}, as well as to reduce system costs or improve power efficiency~\cite{venkataramani2013quality}. 
At the same time, working on floating-point to fixed-point conversion tools allow us to trade-off the algorithm exactness for a more efficient implementation. 
New data types are emerging, such as BFloat16 and Posits.
Recent works on data-type conversion tools~\cite{Cherubin2019ESL,Cattaneo2021DAC} proved ability to adapt to different, heterogeneous architectures.
Optimising the code manually, as done in many embedded systems, is unfeasible in situations where the developer is an expert in the domain, but not in embedded systems software, which is often the case for AI-oriented scenarios. 

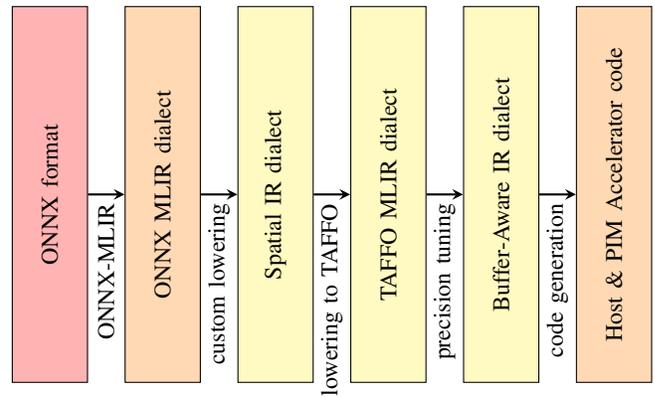
\begin{figure}
    \centering
    \small
\begin{tikzpicture}[node distance=1.5cm]

\node (start) [startstop] {ONNX format};
\node (mlir) [process, below of=start] {ONNX MLIR dialect};
\node (tensor) [quantization, below of=mlir] {Spatial IR dialect};
\node (quantization) [quantization, below of=tensor] {TAFFO MLIR dialect};
\node (buffer) [quantization, below of=quantization] {Buffer-Aware IR dialect};
\node (end) [process, below of=buffer] {Host \& PIM Accelerator code};

\draw [arrow] (start) -- (mlir) node[midway, left, rotate=90] {ONNX-MLIR};
\draw [arrow] (mlir) -- (tensor) node[midway, left, rotate=90] {custom lowering};
\draw [arrow] (tensor) -- (quantization) node[midway, left, rotate=90] {lowering to TAFFO};
\draw [arrow] (quantization) -- (buffer) node[midway, left, rotate=90] {precision tuning};
\draw [arrow] (buffer) -- (end) node[midway, left, rotate=90] {code generation};

\end{tikzpicture}
    \caption{ARCHYTAS Compiler Toolchain: orange boxes represent external dialects, yellow box represent custom dialects capturing the specific aspects of PIM devices}
    \label{fig:compiler}
\end{figure}

In particular, ARCHYTAS will focus on precision tuning techniques automatically applied by the compiler~\cite{cherubin2020tools} with the aim to improve time and energy performance.
%The precision tuning approach will enable us to lower the result accuracy level in exchange for improved non-functional characteristics of the computing system, such as its time and energy performance. 
The ARCHYTAS precision tuner will be based on the state-of-the-art TAFFO precision tuning framework~\cite{Cherubin2020TACO,cattaneo2022taffo} to support the non-conventional computer architectures described in this proposal.
The TAFFO precision tuning framework operates at the compiler level and is built as a modular set of standalone plugins for the open-source industry-grade compiler framework LLVM~\cite{lattner2004llvm}.
This characteristic allows it to easily integrate with other LLVM-based solutions, which are very popular both in industrial and academic contexts.
Starting from programmer hints (attributes), TAFFO performs value range analysis, data type and code conversion, and static estimation of the performance impact. 
TAFFO also provides a fine-grained approach to precision tuning, which makes it easier to be adapted for non-conventional computer architectures.
To better support neural network deployment, ARCHYTAS aims at revising TAFFO to work instead on higher abstractions, via the LLVM-based Multi-Layer Intermediate Representation (MLIR)~\cite{lattner2021mlir}. MLIR has been designed for domain-specific language compilation, and supports directly AI workloads through ONNX-MLIR~\cite{jin2020compiling}, a front-end for the Open Neural Network Exchange (ONNX) model format.
Figure~\ref{fig:compiler} shows the proposed pipeline, relying on custom MLIR dialects to convey information about the geometry of the in-memory-computing solution, as well as a custom TAFFO dialect to support the precision tuning phase.

\section{Conclusions}
\label{sec:conc}
This paper presented the ARCHYTAS EDF project, which aims to investigate and study the feasibility of non-conventional AI accelerators for defense applications that take advantage of novel technologies at the device and package level, such as optoelectronic-based accelerators, volatile and non-volatile processing-in-memory, and neuromorphic devices. The project, driven by a wide consortium of industrial and academic partners from 10 EU member states plus Norway, addresses the challenges encountered in defense sector use cases by proposing optimized solutions for efficient energy consumption, speed, and cost.

\bibliographystyle{splncs04}
\bibliography{biblio.bib}

\end{document}